\documentclass[english,prb,aps,twocolumn,superscriptaddress,showpacs]{revtex4}
\usepackage[latin1]{inputenc}
\usepackage{graphicx}

\makeatletter

\usepackage{babel}
\makeatother
\begin{document}

\title{Spin dynamics in HTSC cuprates: The singlet--correlated band (or
t-J-V) model and its applications}

\author{T. Mayer}

\affiliation{Physics Institue, University of Zurich, CH-8055 Zurich, Switzerland}

\author{M. Eremin}

\affiliation{Kazan State University, Kazan 420008, Russia}

\author{I. Eremin}

\affiliation{Max-Planck Institut für Physik komplexer Systeme, D-01187 Dresden,
Germany}

\affiliation{Technische Universität Braunschweig, D-38106 Braunschweig, Germany}

\author{P. F. Meier}

\affiliation{Physics Institue, University of Zurich, CH-8055 Zurich, Switzerland}

\begin{abstract}
So far calculations of the spin susceptibility in the superconducting
state of cuprates have been performed in the framework of weak--coupling
approximations. However, it is known that cuprates belong to Mott--Hubbard
doped materials where electron correlations are important. In this
paper an analytical expression for the spin susceptibility in the
superconducting state of cuprates is derived within the singlet--correlated
band model, which takes into account strong correlations. The expression
of the spin susceptibility is evaluated using values for the hopping
parameters adapted to measurements of the Fermi surface of the materials
YBa$_{2}$Cu$_{3}$O$_{7}$ and Bi$_{2}$Sr$_{2}$CaCu$_{2}$O$_{8}$.
We show that the available experimental data which are directly related
to the spin susceptibility can be explained consistently within one
set of model parameters for each material. These experiments include
the magnetic resonance peak observed by inelastic neutron scattering
and the temperature dependence of the NMR spin shift, spin--spin and
spin--lattice relaxation rates in the superconducting state. 
\end{abstract}


\maketitle

\section{Introduction}

Derivations of theoretical expressions for the dynamical spin susceptibility
of layered cuprates have been the focus of many ongoing investigations,
since various experimental quantities are directly related to the
spin susceptibility. Large amount of data sets exist on the temperature
dependence of the Knight shift, spin--spin and spin--lattice relaxation
rates as well as inelastic neutron scattering (INS) measurements.
The most complete set of experimental data has been obtained for the
YBaCuO compounds.

The theoretical approaches to the spin susceptibility can be divided
into two categories, the weak-- and strong--coupling models. The former
deals with a single--band Hubbard model with the effective Coulomb
interaction $U_{e}$ taken to be of the order of the bandwidth in
cuprates. Based on this assumption, the dynamical spin susceptibility
can be calculated within a standard random phase approximation (RPA)
approach. Extensive studies of NMR data were carried out within the
framework of this model by Bulut and Scalapino et al.\cite{bulut-scalapino1,bulut-scalapino2,bulut-scalapino3,bulut-scalapino4,bulut-scalapino5}
and Mack et al.\cite{mack}. In the meantime, the features observed
by inelastic neutron scattering were studied theoretically by several
groups\cite{morr-pines,onufrieva-pfeuty}. However, in all of these
calculations the effect of strong electronic correlations were not
taken into account properly. 

At the same time, there were a number of studies devoted to analyzing
the spin susceptibility within the strong--coupling t--J model, for
which standard many--body perturbative methods do not work. For example
the dynamical spin susceptibility was analyzed within the slave--boson
approximation\cite{weng-sheng-ting}, Mori--Zwanzig memory function
formalism\cite{sega-prelovsek-bonca} and with the Hubbard X--operators
technique\cite{zavidonov-brinkmann,eremin-eremin-varlamov}. It has
been found that to a large extent, both weak and strong--coupling
calculations give very similar results for the spin susceptibility.
The respective parameter values, however, differ drastically. Until
now, there is no complete understanding whether both INS and NMR data
can be explained consistently within one model and using the same
parameter values of the given theory. 

In the present work we analyze this question in detail. Starting from
the singlet--correlated band model, we use a well established decoupling
procedure of the equations of motion and approximate higher order
correlation functions so as to obtain an analytical expression for
the dynamical spin susceptibility which takes into account strong
correlations. It has been found by Hubbard and Jain\cite{hubbard-jain}
that the expression for the dynamical susceptibility has a non--trivial
correction due to strong correlation effects. Later Zavidonov and
Brinkmann\cite{zavidonov-brinkmann} have incorporated an additional
functional correction for the lower Hubbard sub--band (LHB) model,
which accounts for local spin fluctuation effects. Both of these corrections
are different from the conventional Pauli--Lindhard form and therefore
they cannot be exactly included in the RPA approach.

In the present paper we extend the previous analysis and present an
analytical expression for the dynamical spin susceptibility in the
upper Hubbard sub--band (UHB) in the superconducting state. We perform
an extended numerical evaluation of this analytical expression and
find that most of the available experimental data which are directly
related to the spin susceptibility can be explained consistently within
one set of model parameters. These experiments include the magnetic
resonance peak observed by INS and the temperature dependence of the
NMR spin shift, spin--spin and the spin--lattice relaxation rates,
measured in the superconducting state. Note that in our analysis we
restrict ourselves to optimally doped high--temperature superconductors,
since the pseudogap phenomenon cannot be explained within our model.
Furthermore, in our analysis we take advantage of other available
experiments, like the Fermi surface topology in various cuprate superconductors
that is determined by high--resolution angle--resolved photoemission.
Assuming a d$_{x^{2}-y^{2}}$--wave pairing symmetry, we propose an
optimal set of parameters for the YBa$_{2}$Cu$_{3}$O$_{7}$ and
Bi$_{2}$Sr$_{2}$CaCu$_{2}$O$_{8}$ compounds. 

This paper is organized as follows. In Sec.\ \ref{sec:Dynamical-spin-susceptiblity}
we introduce the model system and present the analytical expression
for the spin susceptibility in the superconducting state of cuprates.
In Sec.\ \ref{sec:Neutron-scattering-analysis} and Sec.\ \ref{sec:NMR-analysis}
we study the spin susceptibility in the singlet--correlated band model
by analyzing experiments in the superconducting state of the materials
YBa$_{2}$Cu$_{3}$O$_{7}$ and Bi$_{2}$Sr$_{2}$CaCu$_{2}$O$_{8}$.
Summary and conclusions are given in Sec.\ \ref{sec:Conclusions}.

\section{Dynamical spin susceptibility in the singlet--correlated band model\label{sec:Dynamical-spin-susceptiblity}}

The starting point for our calculation is the model Hamiltonian\cite{eremin-eremin-varlamov,eremin-kamaev-eremin} 

\begin{eqnarray}
H & = & \sum_{i,j,\sigma}t_{ij}\psi_{i}^{pd,\sigma}\psi_{j}^{\sigma,pd}+\sum_{i>j}J_{ij}\left[\left(\mathbf{S}_{i}\mathbf{S}_{j}\right)-\frac{n_{i}n_{j}}{4}\right]\label{eq:starting hamiltonian}\\
 & + & \sum_{i>j}V_{ij}\delta_{i}\delta_{j},\nonumber \end{eqnarray}

where $\psi_{i}^{pd,\sigma}$$\left(\psi_{i}^{\sigma,pd}\right)$
are composite copper--oxygen creation (annihilation) operators of
the copper--oxygen singlet\cite{zhang rice} states in the CuO$_{2}$--plane.
Furthermore $J_{ij}$ is the superexchange parameter of the copper
spins (this coupling originates from the virtual hopping from LHB
to UHB via the oxygen state). The number of doped holes is described
by $\delta_{i}=\psi_{i}^{pd,pd}$ and $V_{ij}$ is an effective density--density
interaction parameter. This parameter allows to account for the screened
Coulomb repulsion and phonon (or plasmon) mediated interactions as
well. Consequently, the last term is a Coulomb--like interaction between
doped holes, which can be neglected, because it does not contribute
to the spin susceptibility. 

The susceptibility is calculated from the general expression\begin{equation}
\chi^{+-}\left(\mathbf{q},\omega\right)=-2\pi i\left\langle \left\langle S_{\mathbf{q}}^{+}\left|S_{-\mathbf{q}}^{-}\right.\right\rangle \right\rangle ,\label{eq:susceptibility general expression}\end{equation}

where the spin density operator $S_{\mathbf{q}}^{+}$ for the singlet--correlated
band is written as\begin{eqnarray}
S_{\mathbf{q}}^{+} & = & \sum_{\mathbf{k}}\left(\psi_{\mathbf{k}}^{\uparrow,0}+\psi_{\mathbf{k}}^{pd,\downarrow}\right)\left(\psi_{\mathbf{k}+\mathbf{q}}^{0,\downarrow}-\psi_{\mathbf{k}+\mathbf{q}}^{\uparrow,pd}\right)\label{eq:spinek}\\
 &  & \simeq-\sum_{\mathbf{k}}\psi_{\mathbf{k}}^{pd,\downarrow}\psi_{\mathbf{k}+\mathbf{q}}^{\uparrow,pd}.\nonumber \end{eqnarray}

Here we neglected all the quasiparticle creation (annihilation) operators
$\psi_{i}^{0,\sigma}$$\left(\psi_{i}^{\sigma,0}\right)$ corresponding
to the LHB, because this band is assumed to be completely filled. 

The expression for the susceptibility is derived by the following
procedure. First, we write down a complete set of equations of motion
using the composite copper--oxygen creation (annihilation) operators
$\psi_{i}^{pd,\sigma}$$\left(\psi_{i}^{\sigma,pd}\right)$ of the
copper--oxygen singlet states in the plane. Then, by means of a linear
transformation we rearrange these equations via Bogoliubov's quasiparticle
operators into new sets of equations, which finally will be solved.
An expression for the susceptibility was previously derived\cite{eremin-kamaev-eremin}
by utilizing the method of Heisenberg equations of motion in a small
magnetic field. The advantage of the Green's function method is that
it allows to obtain a formula for the susceptibility which contains
both the itinerant (or quasi Fermi--liquid) part and the local spin
fluctuation part in one general expression.

The equation of motion for the relevant Green's function in the normal
state (T$>$T$_{c}$) has been derived before by some of us\cite{eremin-eremin-varlamov}.
It is given by

\begin{eqnarray}
 &  & \omega\left\langle \left\langle -\psi_{\mathbf{k}}^{pd,\downarrow}\psi_{\mathbf{k}+\mathbf{q}}^{\uparrow,pd}\left|S_{-\mathbf{q}}^{-}\right.\right\rangle \right\rangle =\label{eq:green function 1}\\
 &  & \qquad\qquad\frac{i}{2\pi}\left(\left\langle \psi_{\mathbf{k}}^{pd,\downarrow}\psi_{\mathbf{k}}^{\downarrow,pd}\right\rangle -\left\langle \psi_{\mathbf{k}+\mathbf{q}}^{pd,\uparrow}\psi_{\mathbf{k}+\mathbf{q}}^{\uparrow,pd}\right\rangle \right)\nonumber \\
 &  & \qquad\qquad-\left(\varepsilon_{\mathbf{k}}-\varepsilon_{\mathbf{k}+\mathbf{q}}\right)\left\langle \left\langle -\psi_{\mathbf{k}}^{pd,\downarrow}\psi_{\mathbf{k}+\mathbf{q}}^{\uparrow,pd}\left|S_{-\mathbf{q}}^{-}\right.\right\rangle \right\rangle \nonumber \\
 &  & \qquad\qquad+\frac{1}{N}\left\{ \left(J_{\mathbf{q}}-t_{\mathbf{k}}\right)\left\langle \psi_{\mathbf{k}}^{\downarrow,pd}\psi_{\mathbf{k}}^{pd,\downarrow}\right\rangle \right.\nonumber \\
 &  & \qquad\qquad\left.-\left(J_{\mathbf{q}}-t_{\mathbf{k}+\mathbf{q}}\right)\left\langle \psi_{\mathbf{k+q}}^{pd,\uparrow}\psi_{\mathbf{k+q}}^{\uparrow,pd}\right\rangle \right\} \left\langle \left\langle S_{\mathbf{q}}^{+}\left|S_{-\mathbf{q}}^{-}\right.\right\rangle \right\rangle \nonumber \\
 &  & \qquad\qquad+\frac{P}{N}\sum_{\mathbf{k}'}\left(t_{\mathbf{k}'+\mathbf{q}}-t_{\mathbf{k}'}\right)\left\langle \left\langle \psi_{\mathbf{k}'}^{pd,\downarrow}\psi_{\mathbf{k}'+\mathbf{q}}^{\uparrow,pd}\left|S_{-\mathbf{q}}^{-}\right.\right\rangle \right\rangle ,\nonumber \end{eqnarray}

where the factor $P=(1+\delta)/2$ is a doping dependent constant
which arises due to the narrowing of the band in the so--called Hubbard--I
approximation.

In addition to Eq.\ (\ref{eq:green function 1}) it has been shown\cite{eremin-eremin-varlamov}
that\begin{eqnarray}
 &  & \omega\left\langle \left\langle S_{\mathbf{q}}^{+}\left|S_{-\mathbf{q}}^{-}\right.\right\rangle \right\rangle =\label{eq:green function 2}\\
 &  & \qquad\qquad\sum_{\mathbf{k}'}\left(t_{\mathbf{k}'}-t_{\mathbf{k}'+\mathbf{q}}\right)\left\langle \left\langle \psi_{\mathbf{k'}}^{pd,\downarrow}\psi_{\mathbf{k'}+\mathbf{q}}^{\uparrow,pd}\left|S_{-\mathbf{q}}^{-}\right.\right\rangle \right\rangle .\nonumber \end{eqnarray}

Therefore, if we combine Eq. (\ref{eq:green function 1}) and Eq.
(\ref{eq:green function 2}) we get

\begin{eqnarray}
 &  & \omega\left\langle \left\langle -\psi_{\mathbf{k}}^{pd,\downarrow}\psi_{\mathbf{k}+\mathbf{q}}^{\uparrow,pd}\left|S_{-\mathbf{q}}^{-}\right.\right\rangle \right\rangle =\label{eq:green function 3}\\
 &  & \qquad\qquad\frac{i}{2\pi}\left(\left\langle \psi_{\mathbf{k}}^{pd,\downarrow}\psi_{\mathbf{k}}^{\downarrow,pd}\right\rangle -\left\langle \psi_{\mathbf{k}+\mathbf{q}}^{pd,\uparrow}\psi_{\mathbf{k}+\mathbf{q}}^{\uparrow,pd}\right\rangle \right)\nonumber \\
 &  & \qquad\qquad-\left(\varepsilon_{\mathbf{k}}-\varepsilon_{\mathbf{k}+\mathbf{q}}\right)\left\langle \left\langle -\psi_{\mathbf{k}}^{pd,\downarrow}\psi_{\mathbf{k}+\mathbf{q}}^{\uparrow,pd}\left|S_{-\mathbf{q}}^{-}\right.\right\rangle \right\rangle \nonumber \\
 &  & \qquad\qquad+\frac{1}{N}\left\{ \left(J_{\mathbf{q}}-t_{\mathbf{k}}\right)\left\langle \psi_{\mathbf{k}}^{\downarrow,pd}\psi_{\mathbf{k}}^{pd,\downarrow}\right\rangle \right.\nonumber \\
 &  & \qquad\qquad\left.-\left(J_{\mathbf{q}}-t_{\mathbf{k}+\mathbf{q}}\right)\left\langle \psi_{\mathbf{k+q}}^{pd,\uparrow}\psi_{\mathbf{k+q}}^{\uparrow,pd}\right\rangle \right\} \left\langle \left\langle S_{\mathbf{q}}^{+}\left|S_{-\mathbf{q}}^{-}\right.\right\rangle \right\rangle \nonumber \\
 &  & \qquad\qquad-\frac{P}{N}\omega\left\langle \left\langle S_{\mathbf{q}}^{+}\left|S_{-\mathbf{q}}^{-}\right.\right\rangle \right\rangle .\nonumber \end{eqnarray}

The equation of motion (\ref{eq:green function 3}) allows to derive
the expression of the dynamical spin susceptibility in the normal
state\cite{eremin-eremin-varlamov}.

For the superconducting state we need to perform Bogoliubov's transformation 

\begin{eqnarray}
\alpha_{\mathbf{k}}^{pd,\downarrow} & = & u_{\mathbf{k}}\psi_{\mathbf{k}}^{pd,\downarrow}-v_{\mathbf{k}}\psi_{-\mathbf{k}}^{\uparrow,pd}\label{eq:Bogoliubov transformation}\\
\alpha_{\mathbf{k}}^{pd,\uparrow} & = & u_{\mathbf{k}}\psi_{\mathbf{k}}^{pd,\uparrow}+v_{\mathbf{k}}\psi_{-\mathbf{k}}^{\downarrow,pd}.\nonumber \end{eqnarray}

Consequently, the spin operator for the superconducting state will
be written as

\begin{eqnarray}
S_{\mathbf{q}}^{+} & =\label{eq:Bogoliubov spin}\\
 & - & \sum_{\mathbf{k}}\left(u_{\mathbf{k+q}}u_{\mathbf{k}}\alpha_{\mathbf{k}}^{pd,\downarrow}\alpha_{\mathbf{k}+\mathbf{q}}^{\uparrow,pd}+v_{\mathbf{k}}u_{\mathbf{k}+\mathbf{q}}\alpha_{-\mathbf{k}}^{\uparrow,pd}\alpha_{\mathbf{k}+\mathbf{q}}^{\uparrow,pd}\right)\nonumber \\
 & + & \sum_{\mathbf{k}}\left(u_{\mathbf{k}}v_{\mathbf{k}+\mathbf{q}}\alpha_{\mathbf{k}}^{pd,\downarrow}\alpha_{-\mathbf{k}-\mathbf{q}}^{pd,\downarrow}+v_{\mathbf{k}}v_{\mathbf{k}+\mathbf{q}}\alpha_{-\mathbf{k}}^{\uparrow,pd}\alpha_{-\mathbf{k}-\mathbf{q}}^{pd,\downarrow}\right).\nonumber \end{eqnarray}

Therefore in the superconducting state we need to construct additional
equations for the Green's functions $\left\langle \left\langle -\alpha_{\mathbf{k}}^{pd,\downarrow}\alpha_{\mathbf{k}+\mathbf{q}}^{\uparrow,pd}\left|S_{-\mathbf{q}}^{-}\right.\right\rangle \right\rangle $,
$\left\langle \left\langle -\alpha_{-\mathbf{k}}^{\uparrow,pd}\alpha_{\mathbf{k}+\mathbf{q}}^{\uparrow,pd}\left|S_{-\mathbf{q}}^{-}\right.\right\rangle \right\rangle $,
$\left\langle \left\langle \alpha_{\mathbf{k}}^{pd,\downarrow}\alpha_{-\mathbf{k}-\mathbf{q}}^{pd,\downarrow}\left|S_{-\mathbf{q}}^{-}\right.\right\rangle \right\rangle $
and $\left\langle \left\langle \alpha_{-\mathbf{k}}^{\uparrow,pd}\alpha_{-\mathbf{k}-\mathbf{q}}^{pd,\downarrow}\left|S_{-\mathbf{q}}^{-}\right.\right\rangle \right\rangle $.

Each of them has to be expressed via the $\psi_{\mathbf{k}}^{pd,\sigma}$
operators, for example

\begin{eqnarray}
 &  & \left\langle \left\langle -\alpha_{\mathbf{k}}^{pd,\downarrow}\alpha_{\mathbf{k}+\mathbf{q}}^{\uparrow,pd}\left|S_{-\mathbf{q}}^{-}\right.\right\rangle \right\rangle =\label{eq:express}\\
 &  & \qquad\qquad-u_{\mathbf{k}}u_{\mathbf{k}+\mathbf{q}}\left\langle \left\langle \psi_{\mathbf{k}}^{pd,\downarrow}\psi_{\mathbf{k+q}}^{\uparrow,pd}\left|S_{-\mathbf{q}}^{-}\right.\right\rangle \right\rangle \nonumber \\
 &  & \qquad\qquad+v_{\mathbf{k}}u_{\mathbf{k}+\mathbf{q}}\left\langle \left\langle \psi_{-\mathbf{k}}^{\uparrow,pd}\psi_{\mathbf{k+q}}^{\uparrow,pd}\left|S_{-\mathbf{q}}^{-}\right.\right\rangle \right\rangle \nonumber \\
 &  & \qquad\qquad-u_{\mathbf{k}}v_{\mathbf{k}+\mathbf{q}}\left\langle \left\langle \psi_{\mathbf{k}}^{pd,\downarrow}\psi_{-\mathbf{k-q}}^{pd,\downarrow}\left|S_{-\mathbf{q}}^{-}\right.\right\rangle \right\rangle \nonumber \\
 &  & \qquad\qquad+v_{\mathbf{k}}v_{\mathbf{k}+\mathbf{q}}\left\langle \left\langle \psi_{-\mathbf{k}}^{\uparrow,pd}\psi_{-\mathbf{k-q}}^{pd,\downarrow}\left|S_{-\mathbf{q}}^{-}\right.\right\rangle \right\rangle .\nonumber \end{eqnarray}

Doing so we get

\begin{eqnarray}
 &  & \omega\left\langle \left\langle -\psi_{\mathbf{k}}^{pd,\downarrow}\psi_{\mathbf{k}+\mathbf{q}}^{\uparrow,pd}\left|S_{-\mathbf{q}}^{-}\right.\right\rangle \right\rangle =\label{eq:green function 4}\\
 &  & \qquad\qquad\frac{i}{2\pi}\left(\left\langle \psi_{\mathbf{k}}^{pd,\downarrow}\psi_{\mathbf{k}}^{\downarrow,pd}\right\rangle -\left\langle \psi_{\mathbf{k}+\mathbf{q}}^{pd,\uparrow}\psi_{\mathbf{k}+\mathbf{q}}^{\uparrow,pd}\right\rangle \right)\nonumber \\
 &  & \qquad\qquad+\left\langle \left\langle \left[-\psi_{\mathbf{k}}^{pd,\downarrow}\psi_{\mathbf{k}+\mathbf{q}}^{\uparrow,pd},H\right]\left|S_{-\mathbf{q}}^{-}\right.\right\rangle \right\rangle _{tr}\nonumber \\
 &  & \qquad\qquad+\frac{1}{N}\left\{ \left(J_{\mathbf{q}}-t_{\mathbf{k}}\right)\left\langle \psi_{\mathbf{k}}^{\downarrow,pd}\psi_{\mathbf{k}}^{pd,\downarrow}\right\rangle \right.\nonumber \\
 &  & \qquad\qquad\left.-\left(J_{\mathbf{q}}-t_{\mathbf{k}+\mathbf{q}}\right)\left\langle \psi_{\mathbf{k+q}}^{pd,\uparrow}\psi_{\mathbf{k+q}}^{\uparrow,pd}\right\rangle \right\} \left\langle \left\langle S_{\mathbf{q}}^{+}\left|S_{-\mathbf{q}}^{-}\right.\right\rangle \right\rangle \nonumber \\
 &  & \qquad\qquad-\frac{P}{N}\omega\left\langle \left\langle S_{\mathbf{q}}^{+}\left|S_{-\mathbf{q}}^{-}\right.\right\rangle \right\rangle ,\nonumber \end{eqnarray}

and similar expressions can be obtained for $\left\langle \left\langle \psi_{-\mathbf{k}}^{\uparrow,pd}\psi_{\mathbf{k}+\mathbf{q}}^{\uparrow,pd}\left|S_{-\mathbf{q}}^{-}\right.\right\rangle \right\rangle $,
$\left\langle \left\langle \psi_{\mathbf{k}}^{pd,\downarrow}\psi_{-\mathbf{k}-\mathbf{q}}^{pd,\downarrow}\left|S_{-\mathbf{q}}^{-}\right.\right\rangle \right\rangle $
and $\left\langle \left\langle \psi_{-\mathbf{k}}^{\uparrow,pd}\psi_{-\mathbf{k}-\mathbf{q}}^{pd,\downarrow}\left|S_{-\mathbf{q}}^{-}\right.\right\rangle \right\rangle $.
Equation (\ref{eq:green function 4}) has the same form as in the
normal state (Eq.\ (\ref{eq:green function 3})), except that it
is now adapted to be applied for the superconducting state. We note
that in the conventional Fermi--liquid theory the anticommutator rule
is given as $c_{\mathbf{k}\sigma}c_{\mathbf{k}\sigma}^{\dagger}+c_{\mathbf{k}\sigma}^{\dagger}c_{\mathbf{k}\sigma}=1$.
In the strong--coupling limit, however, this rule is modified\cite{eremin-solovyanov-varlamov}
due to the Coulomb repulsion. For this reason we have abbreviated
the terms which are present in the conventional weak--coupling Fermi--liquid
approach in the superconducting state by the truncated Green's function
$\left\langle \left\langle \left[-\psi_{\mathbf{k}}^{pd,\downarrow}\psi_{\mathbf{k}+\mathbf{q}}^{\uparrow,pd},H\right]\left|S_{-\mathbf{q}}^{-}\right.\right\rangle \right\rangle _{tr}$.
The other terms on the right hand side of Eq.\ (\ref{eq:green function 4})
are due to the spin modulation $S_{\mathbf{q}}^{+}$. 

With the help of these equations we are able to construct the equations
of motion which are needed to calculate the spin susceptibility in
the superconducting state. The first one is given as

\begin{eqnarray}
 &  & \left(\omega-E_{\mathbf{p}}+E_{\mathbf{k}}\right)\left\langle \left\langle \alpha_{\mathbf{k}}^{pd,\downarrow}\alpha_{\mathbf{p}}^{\uparrow,pd}\left|S_{-\mathbf{q}}^{-}\right.\right\rangle \right\rangle =\label{eq:green function 8}\\
 &  & \qquad\qquad\frac{i}{2\pi}\left(u_{\mathbf{k}}u_{\mathbf{p}}+v_{\mathbf{k}}v_{\mathbf{p}}\right)\left(n_{\mathbf{p}}-n_{\mathbf{k}}\right)\nonumber \\
 &  & \qquad\qquad+\frac{1}{N}\left(u_{\mathbf{k}}u_{\mathbf{p}}+v_{\mathbf{k}}v_{\mathbf{p}}\right)\left\{ \left(J_{\mathbf{q}}-t_{\mathbf{p}}\right)n_{\mathbf{p}}\right.\nonumber \\
 &  & \qquad\qquad\left.-\left(J_{\mathbf{q}}-t_{\mathbf{k}}\right)n_{\mathbf{k}}\right\} \left\langle \left\langle S_{\mathbf{q}}^{+}\left|S_{-\mathbf{q}}^{-}\right.\right\rangle \right\rangle \nonumber \\
 &  & \qquad\qquad+\left(Pu_{\mathbf{k}}u_{\mathbf{p}}+(P-1)v_{\mathbf{k}}v_{\mathbf{p}}\right)\frac{\omega}{N}\left\langle \left\langle S_{\mathbf{q}}^{+}\left|S_{-\mathbf{q}}^{-}\right.\right\rangle \right\rangle ,\nonumber \end{eqnarray}

and similar expressions occur for $\left\langle \left\langle \alpha_{-\mathbf{k}}^{\uparrow,pd}\alpha_{-\mathbf{p}}^{pd,\downarrow}\left|S_{-\mathbf{q}}^{-}\right.\right\rangle \right\rangle $,
$\left\langle \left\langle \alpha_{-\mathbf{k}}^{\uparrow,pd}\alpha_{\mathbf{p}}^{\uparrow,pd}\left|S_{-\mathbf{q}}^{-}\right.\right\rangle \right\rangle $
and $\left\langle \left\langle \alpha_{\mathbf{k}}^{pd,\downarrow}\alpha_{-\mathbf{p}}^{pd,\downarrow}\left|S_{-\mathbf{q}}^{-}\right.\right\rangle \right\rangle $.
Furthermore, $n_{\mathbf{k}}=\left\langle \alpha_{\mathbf{k}}^{pd,\uparrow}\alpha_{\mathbf{k}}^{\uparrow,pd}\right\rangle =\left\langle \alpha_{\mathbf{k}}^{pd,\downarrow}\alpha_{\mathbf{k}}^{\downarrow,pd}\right\rangle $
are the occupation numbers in the superconducting state. We further
make use of the identity\begin{eqnarray}
\left\langle \left\langle S_{\mathbf{q}}^{+}\left|S_{-\mathbf{q}}^{-}\right.\right\rangle \right\rangle  & =\label{eq:identity}\\
 & - & \sum_{\mathbf{k}}u_{\mathbf{k}+\mathbf{q}}u_{\mathbf{k}}\left\langle \left\langle \alpha_{\mathbf{k}}^{pd,\downarrow}\alpha_{\mathbf{k}+\mathbf{q}}^{\uparrow,pd}\left|S_{-\mathbf{q}}^{-}\right.\right\rangle \right\rangle \nonumber \\
 & + & \sum_{\mathbf{k}}v_{\mathbf{k}+\mathbf{q}}u_{\mathbf{k}}\left\langle \left\langle \alpha_{\mathbf{k}}^{pd,\downarrow}\alpha_{-\mathbf{k-q}}^{pd,\downarrow}\left|S_{-\mathbf{q}}^{-}\right.\right\rangle \right\rangle \nonumber \\
 & - & \sum_{\mathbf{k}}u_{\mathbf{k}+\mathbf{q}}v_{\mathbf{k}}\left\langle \left\langle \alpha_{-\mathbf{k}}^{\uparrow,pd}\alpha_{\mathbf{k+q}}^{\uparrow,pd}\left|S_{-\mathbf{q}}^{-}\right.\right\rangle \right\rangle \nonumber \\
 & + & \sum_{\mathbf{k}}v_{\mathbf{k}+\mathbf{q}}v_{\mathbf{k}}\left\langle \left\langle \alpha_{-\mathbf{k}}^{\uparrow,pd}\alpha_{-\mathbf{k-q}}^{pd,\downarrow}\left|S_{-\mathbf{q}}^{-}\right.\right\rangle \right\rangle .\nonumber \end{eqnarray}

With help of this relation the susceptibility is calculated as

\begin{eqnarray}
\chi^{+,-}(\mathbf{q},\omega) & =\label{eq:susceptibility}\\
 &  & \frac{\chi_{0}^{+,-}(\mathbf{q},\omega)}{1+J_{\mathbf{q}}\chi_{0}^{+,-}(\mathbf{q},\omega)+\Pi(\mathbf{q},\omega)+Z(\mathbf{q},\omega)},\nonumber \end{eqnarray}

where the superexchange interaction between the copper spins is $J_{\mathbf{q}}=J_{1}\left(\cos q_{x}+\cos q_{y}\right)$,
with $J_{1}$ being the superexchange interaction parameter between
the nearest neighbour copper spins. The function $\chi_{0}^{+,-}(\mathbf{q},\omega)$
is a BCS--like susceptibility and $\Pi(\mathbf{q},\omega)$ is a function
which results from strong correlation effects and has been determined\cite{eremin-kamaev-eremin}
before. It is given by\begin{eqnarray}
\Pi(\mathbf{q},\omega) & = & \frac{P}{N}\sum_{\mathbf{k}}\left(x_{\mathbf{k}}x_{\mathbf{k}+\mathbf{q}}\right.\label{eq:pi function}\\
 &  & \left.+z_{\mathbf{k}}z_{\mathbf{k}+\mathbf{q}}\right)\frac{t_{\mathbf{k}}f_{\mathbf{k}}-t_{\mathbf{k}+\mathbf{q}}f_{\mathbf{k}+\mathbf{q}}}{\omega+i\Gamma+E_{\mathbf{k}}-E_{\mathbf{k}+\mathbf{q}}}\nonumber \\
 & + & \frac{P}{N}\sum_{\mathbf{k}}\left(y_{\mathbf{k}}y_{\mathbf{k}+\mathbf{q}}\right.\nonumber \\
 &  & \left.+z_{\mathbf{k}}z_{\mathbf{k}+\mathbf{q}}\right)\frac{t_{\mathbf{k}}\left(1-f_{\mathbf{k}}\right)-t_{\mathbf{k}+\mathbf{q}}\left(1-f_{\mathbf{k}+\mathbf{q}}\right)}{\omega+i\Gamma-E_{\mathbf{k}}+E_{\mathbf{k}+\mathbf{q}}}\nonumber \\
 & + & \frac{P}{N}\sum_{\mathbf{k}}\left(x_{\mathbf{k}}y_{\mathbf{k}+\mathbf{q}}\right.\nonumber \\
 &  & \left.-z_{\mathbf{k}}z_{\mathbf{k}+\mathbf{q}}\right)\frac{t_{\mathbf{k}}f_{\mathbf{k}}-t_{\mathbf{k}+\mathbf{q}}\left(1-f_{\mathbf{k}+\mathbf{q}}\right)}{\omega+i\Gamma+E_{\mathbf{k}}+E_{\mathbf{k}+\mathbf{q}}}\nonumber \\
 & + & \frac{P}{N}\sum_{\mathbf{k}}\left(y_{\mathbf{k}}x_{\mathbf{k}+\mathbf{q}}\right.\nonumber \\
 &  & \left.-z_{\mathbf{k}}z_{\mathbf{k}+\mathbf{q}}\right)\frac{t_{\mathbf{k}}\left(1-f_{\mathbf{k}}\right)-t_{\mathbf{k}+\mathbf{q}}f_{\mathbf{k}+\mathbf{q}}}{\omega+i\Gamma-E_{\mathbf{k}}-E_{\mathbf{k}+\mathbf{q}}}.\nonumber \end{eqnarray}

The function $Z(\mathbf{q},\omega)$ has its origin in the fast fluctuation
of the localized spins and it is calculated as

\begin{eqnarray}
Z(\mathbf{q},\omega) & = & \frac{1}{N}\sum_{\mathbf{k}}\left(Px_{\mathbf{k}}x_{\mathbf{k}+\mathbf{q}}\right.\label{eq:zet function}\\
 &  & \left.+(P-1)z_{\mathbf{k}}z_{\mathbf{k}+\mathbf{q}}\right)\frac{\omega+i\Gamma}{\omega+i\Gamma+E_{\mathbf{k}}-E_{\mathbf{k}+\mathbf{q}}}\nonumber \\
 & + & \frac{1}{N}\sum_{\mathbf{k}}\left(Py_{\mathbf{k}}y_{\mathbf{k}+\mathbf{q}}\right.\nonumber \\
 &  & \left.+(P-1)z_{\mathbf{k}}z_{\mathbf{k}+\mathbf{q}}\right)\frac{\omega+i\Gamma}{\omega+i\Gamma-E_{\mathbf{k}}+E_{\mathbf{k}+\mathbf{q}}}\nonumber \\
 & + & \frac{1}{N}\sum_{\mathbf{k}}\left(Px_{\mathbf{k}}y_{\mathbf{k}+\mathbf{q}}\right.\nonumber \\
 &  & \left.-(P-1)z_{\mathbf{k}}z_{\mathbf{k}+\mathbf{q}}\right)\frac{\omega+i\Gamma}{\omega+i\Gamma+E_{\mathbf{k}}+E_{\mathbf{k}+\mathbf{q}}}\nonumber \\
 & + & \frac{1}{N}\sum_{\mathbf{k}}\left(Py_{\mathbf{k}}x_{\mathbf{k}+\mathbf{q}}\right.\nonumber \\
 &  & \left.-(P-1)z_{\mathbf{k}}z_{\mathbf{k}+\mathbf{q}}\right)\frac{\omega+i\Gamma}{\omega+i\Gamma-E_{\mathbf{k}}-E_{\mathbf{k}+\mathbf{q}}},\nonumber \end{eqnarray}

where the functions $x_{\mathbf{k}}=u_{\mathbf{k}}^{2}=\frac{1}{2}\left(1+\varepsilon_{\mathbf{k}}/E_{\mathbf{k}}\right)$,
$y_{\mathbf{k}}=v_{\mathbf{k}}^{2}=\frac{1}{2}\left(1-\varepsilon_{\mathbf{k}}/E_{\mathbf{k}}\right)$
and $z_{\mathbf{k}}=u_{\mathbf{k}}v_{\mathbf{k}}=\Delta_{\mathbf{k}}/(2E_{\mathbf{k}})$
are the conventional coherence factors. Furthermore $\Gamma$ is an
artificially introduced damping constant and $E_{\mathbf{k}}=\sqrt{\left(\varepsilon_{\mathbf{k}}-\mu\right)^{2}+\Delta_{\mathbf{k}}^{2}}$
is the energy of Bogoliubov's quasiparticles in the superconducting
state. The energy dispersion in tight--binding approximation for a
quadratic two--dimensional lattice is given as \begin{eqnarray}
\varepsilon_{\mathbf{k}} & = & P\left[2t_{1}\left(\cos k_{x}+\cos k_{y}\right)\right.\label{eq:tight binding}\\
 & + & 4t_{2}\left(\cos k_{x}\cos k_{y}\right)\nonumber \\
 & + & 2t_{3}\left(\cos2k_{x}+\cos2k_{y}\right)\nonumber \\
 & + & 2t_{4}\left(\cos2k_{x}\cos k_{y}+\cos2k_{y}\cos k_{x}\right)\nonumber \\
 & + & \left.4t_{5}\left(\cos2k_{x}\cos2k_{y}\right)\right],\nonumber \end{eqnarray}

where the model parameters $t_{1},t_{2}...$ correspond to nearest--neighbour
(NN), next--nearest--neighbour (NNN), and further distant hopping,
respectively. For simplicity we do not consider hopping between layers.
Further we note that at optimal doping the number of doped holes per
unit cell in one CuO$_{2}$--layer is $0.165$. In bilayer
compounds therefore we have $\delta=0.33$ with a corresponding factor
$P=(1+\delta)/2\simeq0.7$ near optimal doping.

The mechanism that causes the pairing in cuprates is still being debated
and the origin of the interactions described by $V_{ij}$ in Eq.\
(\ref{eq:starting hamiltonian}) are unknown. Therefore we have introduced
the superconducting gap function $\Delta_{\mathbf{k}}$ phenomenologically
into our model. Assuming a d$_{x^{2}-y^{2}}$ pairing symmetry it
is given by

\begin{equation}
\Delta_{\mathbf{k}}(T)=\frac{\Delta_{0}}{2}\left(\cos k_{x}-\cos k_{y}\right)\tanh\left(1.76\sqrt{T_{c}/T-1}\right),\label{eq:gap equation model}\end{equation}

where $\Delta_{0}$ is considered to be a model parameter. We would
like to point out that this formula is a fit to the solution of the
Eliashberg strong--coupling gap equation. %
\begin{figure}
\begin{center}\includegraphics[%
  scale=0.4]{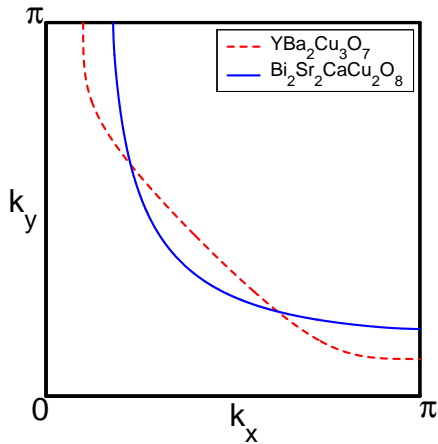}\end{center}

\caption{\label{cap:Fermi-surfaces}Fermi surfaces of YBa$_{2}$Cu$_{3}$O$_{7}$
(dashed) and Bi$_{2}$Sr$_{2}$CaCu$_{2}$O$_{8}$ (solid) according
to photoemission experiments.}
\end{figure}

In the forthcoming sections we will analyze several experiments in
the superconducting state of the materials YBa$_{2}$Cu$_{3}$O$_{7}$
and Bi$_{2}$Sr$_{2}$CaCu$_{2}$O$_{8}$. We would like to summarize
at this point the parameters which we used to perform our analysis.
The tight--binding hopping parameters $\left(Pt_{1}...Pt_{5}\right)$
are adopted from fits to the measured Fermi surfaces of these two
materials. For YBa$_{2}$Cu$_{3}$O$_{7}$ we have\cite{norman} $\left(\mu,Pt_{1}...Pt_{5}\right)=\left(119,147,-36.5,-2.4,32.4,-1.8\right)$~meV,
while for Bi$_{2}$Sr$_{2}$CaCu$_{2}$O$_{8}$ we take\cite{eschrig-norman}
$\left(\mu,Pt_{1}...Pt_{5}\right)=\left(49.4,73.9,-12.0,16.3,6.3,-11.7\right)$~meV.
The corresponding Fermi surfaces are shown in Fig.\ \ref{cap:Fermi-surfaces}.
Note that they are quite different for these two materials. 

Other model parameters include the gap parameter which is assumed
to be in the order of $\Delta_{0}\simeq10-30$~meV and the superexchange
interaction parameter of the copper spins which is $J_{1}\simeq100-140$~meV.
In our analysis we proceed as follows. We assume the Fermi surface
as given from fits to photo\-emission data and thus fix the values
of the hopping parameters for both materials. Based on this assumption
we analyze neutron scattering experiments which allows us to determine
the values of the model parameters $\Delta_{0}$ and $J_{1}$. Then
we move to NMR experiments and calculate the temperature dependences
of spin shift, spin--spin relaxation and spin--lattice relaxation
rates, utilizing the parameter values obtained before. Thus in our
analysis we try to establish a connection between the different experiments.

\section{Neutron scattering analysis\label{sec:Neutron-scattering-analysis}}

Magnetic inelastic neutron scattering experiments probe directly the
imaginary part of the dynamical spin susceptibility $Im\chi^{+,-}\left(\mathbf{q},\omega\right)$.
The experiments indicate a sharp resonance in the magnetic excitation
spectrum of optimally doped YBa$_{2}$Cu$_{3}$O$_{7}$\cite{bourges,mook}
and Bi$_{2}$Sr$_{2}$CaCu$_{2}$O$_{8}$\cite{fong,he} compounds
at a frequency $\omega\simeq41$~meV, near the antiferromagnetic
wave vector $\mathbf{Q}=\left(\pi,\pi\right)$. Consequently, there
should be a large peak in the imaginary part of the spin susceptibility
$Im\chi^{+,-}\left(\mathbf{\mathbf{Q}},\omega\right)$ at the same
frequency. In the conventional weak--coupling scenario this feature
was studied extensively by various authors\cite{morr-pines,onufrieva-pfeuty},
\begin{figure}
\begin{center}\includegraphics[%
  scale=0.4]{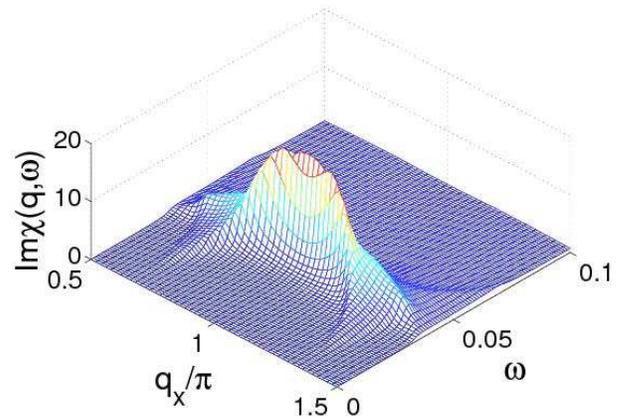}\end{center}

\begin{center}\includegraphics[%
  scale=0.4]{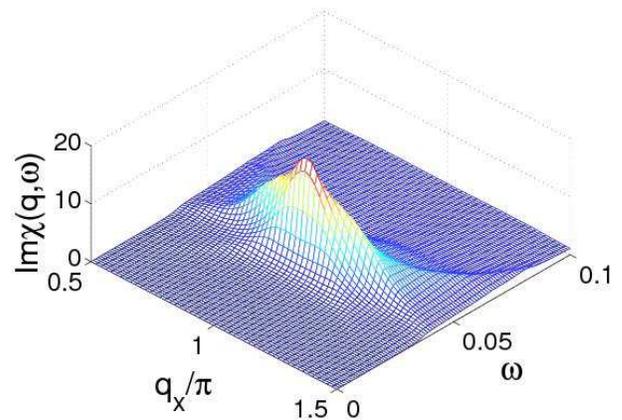}\end{center}

\caption{\label{cap:imchi}Calculated frequency and momentum dependence of
$Im\chi^{+,-}\left(\mathbf{q},\omega\right)$ for YBa$_{2}$Cu$_{3}$O$_{7}$
(top) and Bi$_{2}$Sr$_{2}$CaCu$_{2}$O$_{8}$ (bottom).}
\end{figure}
who connected the appearance of the resonance peak to a collective
spin--density wave mode formation. Special experimental features like
the effect of orthorhombic distortions\cite{hinkov} and bilayer splitting\cite{pailhes}
on the magnetic excitations were also studied theoretically\cite{eremin-manske,eremin private}
within the weak--coupling model. In the strong--coupling limit previous
calculations were carried out by some of us\cite{eremin-kamaev-eremin}
and the dependence of the position of the resonance peak on the model
parameters was studied extensively. We will not repeat these considerations
here. Note only that within our model the position of the neutron
scattering resonance peak is determined mainly by the magnitude of
the superconducting gap $\Delta_{0}$ and the superexchange parameter
$J_{1}$. In particular, the superconducting gap parameter $\Delta_{0}$
determines the size of the transparency window in $Im\chi^{+,-}\left(\mathbf{q},\omega\right)$
which is approximately $\omega\simeq2\Delta_{0}$. In this region
a sharp delta--like peak appears in the imaginary part of the susceptibility
if the resonance condition $1+J_{\mathbf{q}}Re\chi_{0}^{+,-}(\mathbf{q},\omega)+Re\Pi(\mathbf{q},\omega)+ReZ(\mathbf{q},\omega)=0$
(see Eq.\ (\ref{eq:susceptibility})) is fulfilled. Our calculations
indicate that for YBa$_{2}$Cu$_{3}$O$_{7}$ (Bi$_{2}$Sr$_{2}$CaCu$_{2}$O$_{8}$)
the value of the gap parameter should be $\Delta_{0}=24$~meV ($25$~meV).
The corresponding values of the superexchange parameter are determined
as $J_{1}=90$~meV ($110$~meV). For these values the resonance
condition is fulfilled and a clear peak appears in the imaginary part
of the susceptibility near $\omega\simeq41$~meV, for both materials.
In Fig.\ \ref{cap:imchi} we display the calculated momentum and
frequency dependence of the imaginary part of the susceptibility.
Note that the height of the resonance peak depends on the artificially
introduced quasiparticle damping $\Gamma$, therefore the values of
$Im\chi^{+,-}\left(\mathbf{q},\omega\right)$ in Fig.\ \ref{cap:imchi}
are arbitrary. Furthermore, the experimentally reported\cite{reznik}
downward dispersion branch for YBa$_{2}$Cu$_{3}$O$_{7}$ with respect
to $\omega$ is reproduced by our model calculations as can be seen
from an inspection of Fig.\ \ref{cap:imchi} (top). We would like
to point out that for the Bi$_{2}$Sr$_{2}$CaCu$_{2}$O$_{8}$ compound
we do not find a similar dispersion branch. There are, however, no
experiments available which would allow a comparison.%
\begin{figure}
\begin{center}\includegraphics[%
  scale=0.3]{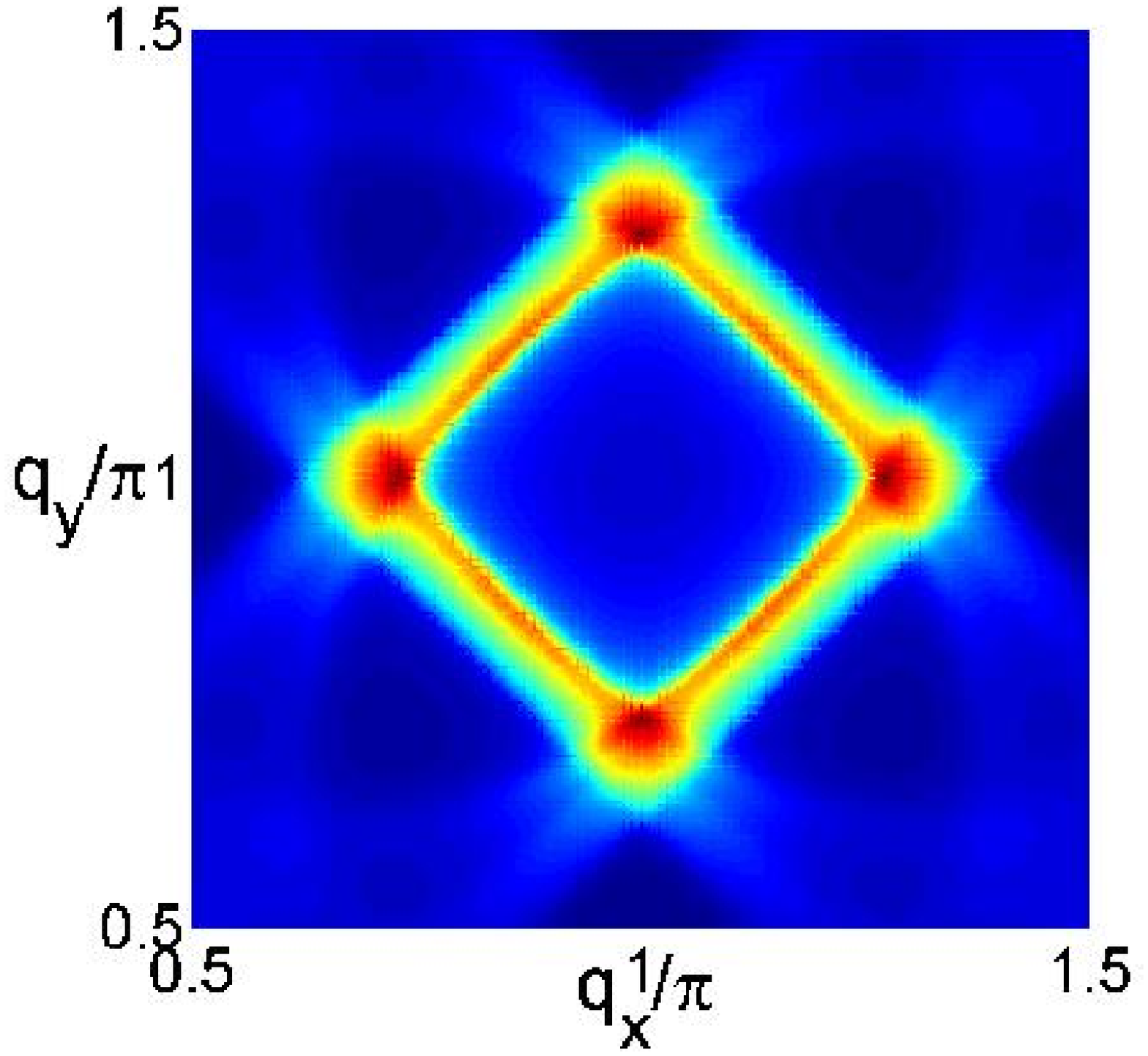}\end{center}

\begin{center}\includegraphics[%
  scale=0.3]{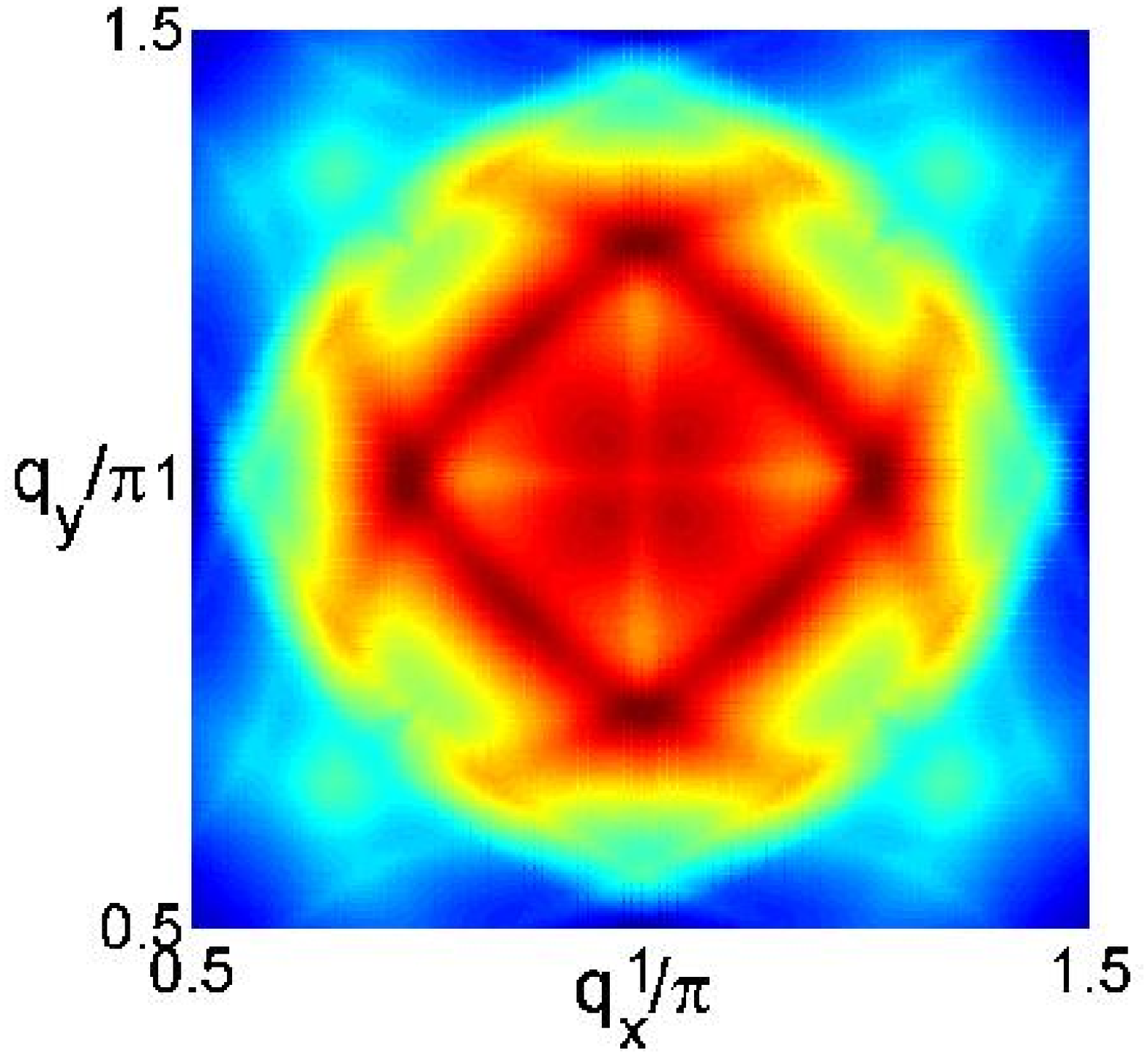}\end{center}

\caption{\label{cap:Low-energy} Intensity plot of the imaginary part of the
susceptibility $Im\chi^{+,-}\left(\mathbf{q},\omega\right)$ near
$\mathbf{Q}$ for YBa$_{2}$Cu$_{3}$O$_{7}$ (top) and Bi$_{2}$Sr$_{2}$CaCu$_{2}$O$_{8}$
(bottom) for $\omega=30$~meV.}
\end{figure}

Let us now turn to the examination of magnetic excitations at lower
frequency, where measurements\cite{mook-pengcheng-hayden-aeppli}
in the YBaCuO compounds indicate well defined incommensurability in
the magnetic excitation spectrum. For the material Bi$_{2}$Sr$_{2}$CaCu$_{2}$O$_{8}$,
however, the data sets\cite{mook--dogan--chakoumakos} are inconclusive
due to experimental difficulty. Here we report significant differences
in the low frequency excitations for the materials YBa$_{2}$Cu$_{3}$O$_{7}$
and Bi$_{2}$Sr$_{2}$CaCu$_{2}$O$_{8}$. Similar conclusions were
reached previously by Norman\cite{normaninc} in the conventional
weak--coupling scenario. In Fig.\ \ref{cap:Low-energy} we show an
intensity plot of the imaginary part of the susceptibility around
the antiferromagnetic wave vector $\mathbf{Q}$ calculated for $\omega=30$~meV,
for both materials. By examination of the figure we conclude that
for YBa$_{2}$Cu$_{3}$O$_{7}$ the model calculations match the experimental
observation\cite{mook-pengcheng-hayden-aeppli} of incommensurability.
For the Bi$_{2}$Sr$_{2}$CaCu$_{2}$O$_{8}$ compound, however, our
results indicate that the incommensurability of the magnetic excitations
is much weaker than in YBa$_{2}$Cu$_{3}$O$_{7}$ due to the difference
in the Fermi surface topology. This result could be tested by further
experiments.

\section{NMR analysis\label{sec:NMR-analysis}}

\subsection{Knight shift}

In order to calculate the Knight shift in the superconducting state
we need to calculate the susceptibility in the limit $\mathbf{q}\rightarrow0$,
$\omega=0$. The BCS susceptibility $\chi_{0}^{+,-}\left(\mathbf{q},\omega\right)$
converts to the Yosida result\cite{yosida} 

\begin{equation}
\chi_{0}^{+,-}\left(\mathbf{q}\rightarrow0,\omega=0\right)\simeq\frac{P\beta}{N}\sum_{\mathbf{k}}\frac{\partial f\left(E_{\mathbf{k}}\right)}{\partial E_{\mathbf{k}}}=\chi_{P}.\label{eq:yosida}\end{equation}

The functions $\Pi(\mathbf{q},\omega)$ and $Z(\mathbf{q},\omega)$
are approximated as

\begin{eqnarray}
\Pi(\mathbf{q}\rightarrow0,\omega=0) & \simeq & \frac{1}{N}\sum_{\mathbf{k}}f\left(E_{\mathbf{k}}\right)\label{eq:approx pi}\\
 &  & -\frac{P\beta}{N}\sum_{\mathbf{k}}t_{\mathbf{k}}\frac{\partial f\left(E_{\mathbf{k}}\right)}{\partial E_{\mathbf{k}}}\nonumber \\
 & \simeq & \frac{\delta}{P}-\frac{P\beta}{N}\sum_{\mathbf{k}}t_{\mathbf{k}}\frac{\partial f\left(E_{\mathbf{k}}\right)}{\partial E_{\mathbf{k}}}\nonumber \end{eqnarray}

and \begin{equation}
Z(\mathbf{q}\rightarrow0,\omega=0)\simeq P.\label{eq:approx zet}\end{equation}

In the long--wave limit therefore the susceptibility is given by the
simple expression\begin{equation}
\chi^{+,-}(\mathbf{q}\rightarrow0,\omega=0)=\frac{\chi_{P}}{1+P+\delta/P+\left(2J_{1}-\mu/P\right)\chi_{P}}.\label{eq:approx chi}\end{equation}

\begin{figure}
\begin{center}\includegraphics[%
  scale=0.4]{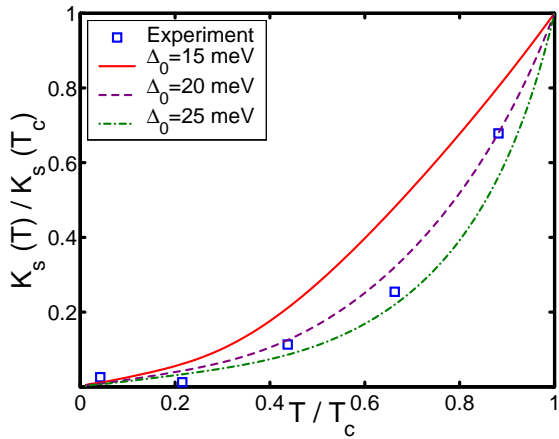}\end{center}

\caption{\label{cap:spin shifts for ybacuo j=3D0}Temperature dependence of
the reduced spin shift in YBa$_{2}$Cu$_{3}$O$_{7}$ for no exchange
interaction ($J_{1}=0$ eV). The experimental points are taken from
Barrett et al.\cite{barrett}.}
\end{figure}

With the help of this relation the spin shift can be calculated according
to 

\begin{equation}
K_{s}\propto A\chi^{+,-}(\mathbf{q}\rightarrow0,\omega=0),\label{eq:knight shift}\end{equation}

where $A$ represents the appropriate hyperfine coupling constant.
Note that this expression refers to the spin contribution to the magnetic
shift. In addition there is an orbital (chemical) shift which, however,
is independent of the temperature. We will calculate the temperature
dependence of the normalized spin shifts $K_{s}(T)/K_{s}(T_{c})$.
In this way the hyperfine coupling constants in Eq.\ (\ref{eq:knight shift})
cancel out, simplifying our analysis. Furthermore, note that the quasiparticle
damping is $\Gamma\rightarrow0^{+}$. In Fig.\ \ref{cap:spin shifts for ybacuo j=3D0}
we display the calculated temperature dependence of the spin shifts
for YBa$_{2}$Cu$_{3}$O$_{7}$, along with the experimental points
of Barrett et al.\cite{barrett}, for vanishing superexchange interaction
$J_{1}=0$~eV and different values of the gap parameter $\Delta_{0}$.
We observe that below T$_{c}$ the spin shifts depend strongly on
the magnitude of the gap parameter $\Delta_{0}$. This behavior has
also been found for the RPA susceptibility\cite{bulut-scalapino5,mack}.

\begin{figure}
\begin{center}\includegraphics[%
  scale=0.4]{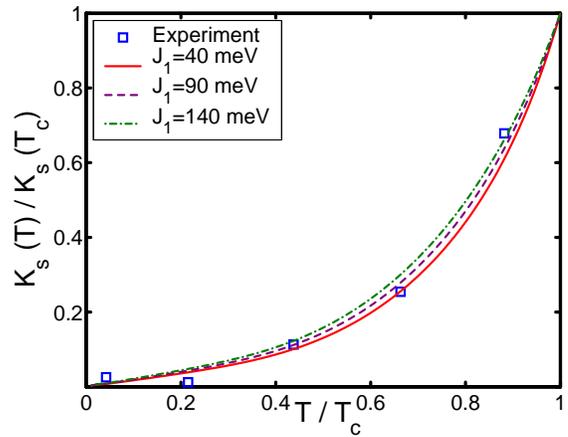}\end{center}

\caption{\label{cap:spin shifts for Ybacuo j}Temperature dependence of the
reduced spin shift in YBa$_{2}$Cu$_{3}$O$_{7}$. The energy gap
is $\Delta_{0}=24$~meV. The experimental points are taken from Barrett
et al.\cite{barrett}.}
\end{figure}

Next we consider how the spin shift depends on the superexchange interaction
$J_{1}$. In Fig.\ \ref{cap:spin shifts for Ybacuo j} the calculated
spin shifts for different values of the superexchange interaction
parameter $J_{1}$ are shown. We see that the temperature dependence
of the Knight shift does not significantly change by adjusting the
parameter $J_{1}$. Also, contrary to the RPA scenario, the superexchange
coupling $J_{1}$ reduces the rapid decrease of the Knight shift.
By analysis of the figure we conclude that the optimal set of parameters
to describe the experimentally observed temperature dependence of
the spin shift for YBa$_{2}$Cu$_{3}$O$_{7}$ is $\Delta_{0}=24$~meV
and $J_{1}=90$~meV for the given Fermi surface. These values are
in perfect agreement with those determined from the fit to neutron
scattering experiments in the previous section.%

\begin{figure}
\begin{center}\includegraphics[%
  scale=0.4]{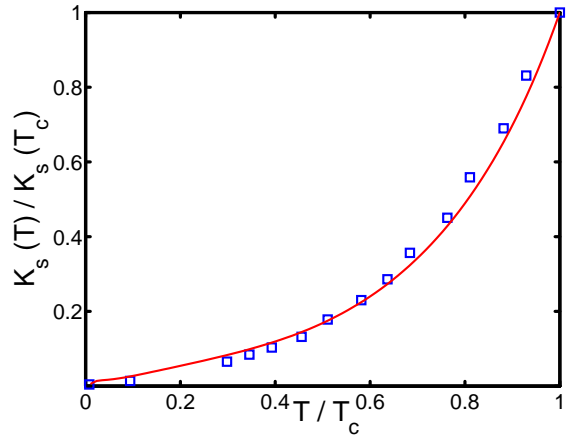}\end{center}

\caption{\label{cap:spin shifts bisco}Temperature dependence of the reduced
spin shift in Bi$_{2}$Sr$_{2}$CaCu$_{2}$O$_{8}$. The experimental
points are taken from Ishida et al.\cite{ishida}.}
\end{figure}

Let us now turn to the examination of the spin shift in the Bi$_{2}$Sr$_{2}$CaCu$_{2}$O$_{8}$
compound. Experimental results indicate a similar behaviour as in
the YBa$_{2}$Cu$_{3}$O$_{7}$ material: the spin shift decreases
rapidly upon entering the superconducting state. The calculated spin
shifts also show a similar dependence on the model parameters $\Delta_{0}$
and $J_{1}$. We will not repeat the analysis of these dependences
and show instead in Fig.\ \ref{cap:spin shifts bisco} the final
result of our calculations for the spin shift in Bi$_{2}$Sr$_{2}$CaCu$_{2}$O$_{8}$
along with the experimental points of Ishida et al.\cite{ishida}.
The parameters used for the calculation are $\Delta_{0}=24$~meV
and $J_{1}=110$~meV. We notice that again these values almost coincide
with those determined by the analysis of neutron scattering experiments
in this compound. By examination of the figure we see that the calculated
temperature dependence of spin shift gives a satisfactory fit to the
experimental data. %

Finally we conclude that it is possible to account for both the neutron
scattering resonance peak and the temperature dependence of the spin
shift in the superconducting state of YBa$_{2}$Cu$_{3}$O$_{7}$
and Bi$_{2}$Sr$_{2}$CaCu$_{2}$O$_{8}$ consistently within the
same set of parameters for each material. Next, we calculate the temperature
dependence of dynamical NMR quantities, the spin--spin and spin--lattice
relaxation rates.

\subsection{Spin--spin relaxation}

The nuclear spin--spin relaxation rate is calculated from the expression\cite{thelen-pines}\begin{eqnarray}
T_{2G}^{-2} & \propto & \left[\frac{1}{N}\sum_{\mathbf{q}}\,^{63}F_{\Vert}(\mathbf{q})^{2}\left(Re\chi^{+,-}(\mathbf{q},\omega=0)\right)^{2}\right.\label{eq:t2g}\\
 &  & \qquad\left.-\left(\frac{1}{N}\sum_{\mathbf{q}}\,^{63}F_{\Vert}(\mathbf{q})Re\chi^{+,-}(\mathbf{q},\omega=0)\right)^{2}\right],\nonumber \end{eqnarray}

where $^{63}F_{\Vert}(\mathbf{q})=\left[A_{\Vert}+2B\left(\cos(q_{x})+\cos(q_{y})\right)\right]^{2}$
is the hyperfine form factor. The values of the hyperfine coupling
constants are taken as $B\simeq0.4$~$\mu$eV and $A_{\Vert}\simeq-4B$.
The spin--spin relaxation rate allows us to study the real part of
the susceptibility $Re\chi^{+,-}(\mathbf{q},\omega=0)$ near the antiferromagnetic
wave vector $\mathbf{Q}$. We will calculate the temperature dependence
of the normalized spin--spin relaxation rate $T_{2G}^{-1}(T)/T_{2G}^{-1}(T_{c})$
in the same way as we analyzed the Knight shifts. Note that for the
evaluation of the spin--spin relaxation, the real part of the susceptibility
is calculated by taking the quasiparticle damping $\Gamma\rightarrow0^{+}$.
Otherwise, due to the behavior of the coherence factors, a large increase
in the spin--spin relaxation rate occurs near T$_{c}$ upon entering
the superconducting state as has been discussed in Ref.\cite{thelen-pines}.
\begin{figure}
\begin{center}\includegraphics[%
  scale=0.4]{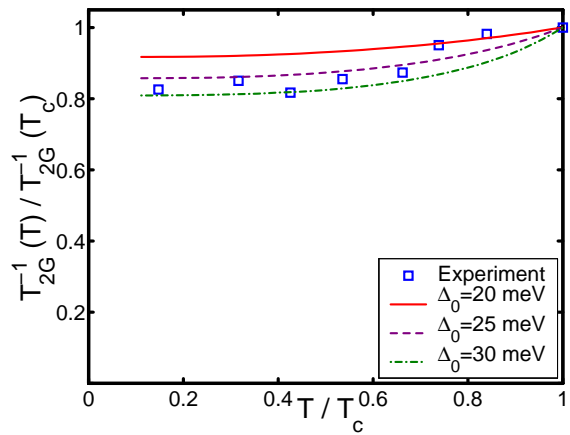}\end{center}

\caption{\label{cap:spin-spin relax 1}Temperature dependence of the spin--spin
relaxation rate in YBa$_{2}$Cu$_{3}$O$_{7}$ for no exchange interaction
($J_{1}=0$~eV). The experimental points are taken from Stern et
al.\cite{stern}.}
\end{figure}

In Fig.\ \ref{cap:spin-spin relax 1} we display the calculated spin--spin
relaxation rates for YBa$_{2}$Cu$_{3}$O$_{7}$ along with experimental
points from Stern et al.\cite{stern} for no superexchange interaction
$J_{1}=0$~eV. We observe that the results show a similar temperature
dependence as in the RPA approach\cite{mack}. Generally, the temperature
dependence of the spin--spin relaxation rate is less sensitive to
the change of the gap parameter than the spin shift. We see that for
the hypothetical case of no interaction we can account for the observed
temperature dependence of the spin--spin relaxation rate. Next we
wish to study the behavior of the spin--spin relaxation for different
values of the superexchange interaction parameter $J_{1}$. 

\begin{figure}
\begin{center}\includegraphics[%
  scale=0.4]{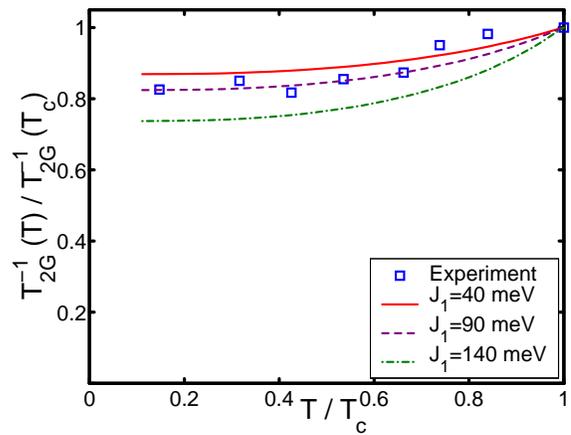}\end{center}

\caption{\label{cap:spin-spin relax 2}Temperature dependence of the spin--spin
relaxation rate in YBa$_{2}$Cu$_{3}$O$_{7}$. The energy gap is
$\Delta_{0}=22$~meV. The experimental points are taken from Stern
et al.\cite{stern}.}
\end{figure}

In Fig.\ \ref{cap:spin-spin relax 2} the temperature dependence
of the spin--spin relaxation rate is shown for various values of the
superexchange interaction parameter $J_{1}$. We see that we get a
reasonable agreement with the data using the parameter values $\Delta_{0}=22$~meV
and $J_{1}=90$~meV. The magnitudes of these parameters are in agreement
with those obtained by the analysis of neutron scattering and NMR
spin shift experiments for YBa$_{2}$Cu$_{3}$O$_{7}$ . %

Concerning the Bi$_{2}$Sr$_{2}$CaCu$_{2}$O$_{8}$ compound our
calculations indicate a very similar behaviour as in YBa$_{2}$Cu$_{3}$O$_{7}$,
if we utilize the parameter values from the previous sections $\Delta_{0}=24$~meV
and $J_{1}=110$~meV. Unfortunately the spin--spin relaxation rate
has not yet been measured in Bi$_{2}$Sr$_{2}$CaCu$_{2}$O$_{8}$
, thus we have no basis for comparison with experiments.

\subsection{Spin--lattice relaxation}

The nuclear spin--lattice relaxation rate is calculated according
to the expression\cite{moriya}

\begin{equation}
^{\alpha}T_{1\beta}^{-1}\propto\frac{T}{N}\sum_{\mathbf{q},\beta'}\,^{\alpha}F_{\beta'}(\mathbf{q})\lim_{\omega\rightarrow0}\frac{Im\chi^{+,-}(\mathbf{q},\omega)}{\omega},\label{eq:t1t}\end{equation}

where $\beta$ denotes the field direction and $\beta'$ are the directions
orthogonal to the field. Furthermore $\alpha$ designates the nucleus
under consideration.

In order to calculate the imaginary part of the spin susceptibility
$Im\chi^{+,-}(\mathbf{q},\omega\rightarrow0)$ we introduced a finite
quasiparticle broadening $\Gamma=3k_{B}T_{c}\simeq2$~meV, following
the analysis of Bulut and Scalapino\cite{bulut-scalapino5}. Furthermore,
the form factors in Eq.\ (\ref{eq:t1t}) are given by

\begin{eqnarray}
^{63}F_{\beta}(\mathbf{q}) & = & \left[A_{\beta}+2B\left(\cos(q_{x})+\cos(q_{y})\right)\right]^{2},\label{eq:form factors}\\
^{17}F_{\beta}(\mathbf{q}) & = & 2\left(C_{\beta_{1}}^{2}\cos^{2}(q_{x}/2)+C_{\beta_{2}}^{2}\cos^{2}(q_{y}/2)\right).\nonumber \end{eqnarray}

The values of the hyperfine coupling constants are taken as $B\simeq0.4$
$\mu$eV, $A_{\Vert}\simeq-4B$, $A_{\bot}\simeq0.75B$, $C_{\Vert}\simeq0.6B$,
and $C_{\bot}\simeq0.32B$. 

\begin{figure}
\begin{center}\includegraphics[%
  scale=0.4]{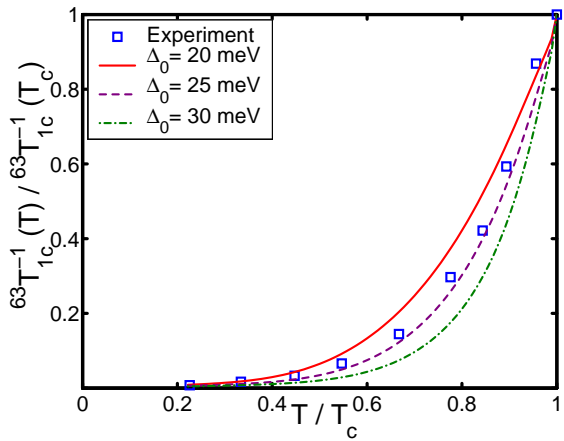}\end{center}

\caption{\label{cap:t1t delta}Temperature dependence of the spin--lattice
relaxation rate in YBa$_{2}$Cu$_{3}$O$_{7}$ for no exchange interaction
($J_{1}=0$~eV). The experimental points are taken from Takigawa
et al.\cite{takigawa}. }
\end{figure}

In Fig.\ \ref{cap:t1t delta} we display the calculated spin--lattice
relaxation rates for YBa$_{2}$Cu$_{3}$O$_{7}$ along with the experimental
points of Takigawa et al.\cite{takigawa}, when the superexchange
interaction is $J_{1}=0$~eV. We observe that the temperature dependence
varies strongly when adjusting the gap parameter $\Delta_{0}$. As
for the spin shift and spin--spin relaxation rate calculations it
is possible to fit the experimental data even without taking into
account the interaction. %

\begin{figure}
\begin{center}\includegraphics[%
  scale=0.4]{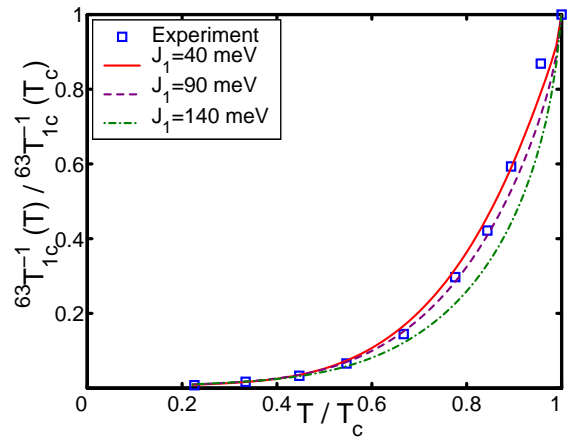}\end{center}

\caption{\label{cap:t1t jot}Temperature dependence of the spin--lattice relaxation
rate in YBa$_{2}$Cu$_{3}$O$_{7}$. The energy gap is $\Delta_{0}=22$~meV.
The experimental points are taken from Takigawa et al.\cite{takigawa}. }
\end{figure}

Next we consider the effect of the superexchange parameter $J_{1}$.
In Fig.~\ref{cap:t1t jot} the spin--lattice relaxation rate is shown
for different values of $J_{1}$. We note that upon changing the values
of $J_{1}$ the spin--lattice relaxation rate $T_{1c}^{-1}$ changes
the same way as it does in the RPA case if the parameter value of
the effective Coulomb interaction $U_{e}$ is changed. Namely, the
parameter $J_{1}$ has no significant impact on the temperature dependence
of the spin--lattice relaxation rate in the superconducting state.
Upon further examination of the figure we see that we get a reasonable
agreement with experimental observation using the parameter values
$\Delta_{0}=22$~meV and $J_{1}=90$~meV. These parameters agree
with those we determined before in the previous Sections. %

\begin{figure}
\begin{center}\includegraphics[%
  scale=0.4]{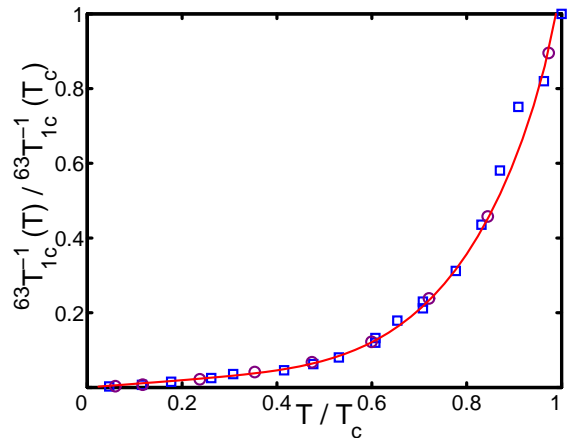}\end{center}

\caption{\label{cap:t1t bisco}Temperature dependence of the spin--lattice
relaxation rate in Bi$_{2}$Sr$_{2}$CaCu$_{2}$O$_{8}$. The experimental
points are taken from Ishida et al.\cite{ishida} (squares) and Takigawa
et al.\cite{takigawa 2} (circles).}
\end{figure}

Next we examine the spin--lattice relaxation rate in the Bi$_{2}$Sr$_{2}$CaCu$_{2}$O$_{8}$
compound. Upon changing the model parameters $\Delta_{0}$ and $J_{1}$
the spin--lattice relaxation behaves much the same way as in YBa$_{2}$Cu$_{3}$O$_{7}$.
We show in Fig.\ \ref{cap:t1t bisco} the final result of our calculations
of the spin--lattice relaxation rate in Bi$_{2}$Sr$_{2}$CaCu$_{2}$O$_{8}$,
along with the experimental points of Ishida et al.\cite{ishida}
(squares) and Takigawa et al.\cite{takigawa 2} (circles). The parameters
used for the calculation are $\Delta_{0}=23$~meV and $J_{1}=110$~meV.
Note that the values of these parameters are again almost the same
as those that we used before when we analyzed neutron scattering and
spin shift experiments. A particularly interesting feature can be
found when comparing the spin--lattice relaxation rates in YBa$_{2}$Cu$_{3}$O$_{7}$
and Bi$_{2}$Sr$_{2}$CaCu$_{2}$O$_{8}$ at low temperatures. A close
inspection of the corresponding figures (Fig.\ \ref{cap:t1t jot}
and Fig.\ \ref{cap:t1t bisco}) shows that in the former case $^{63}T_{1c}^{-1}(T)$
practically vanishes at temperatures T$<20$~K, while in the latter
case the relaxation rate seems to vanish only at T$\simeq0$~K. Note
that both of these dependences are reproduced by the model calculations.%

\begin{figure}
\begin{center}\includegraphics[%
  scale=0.4]{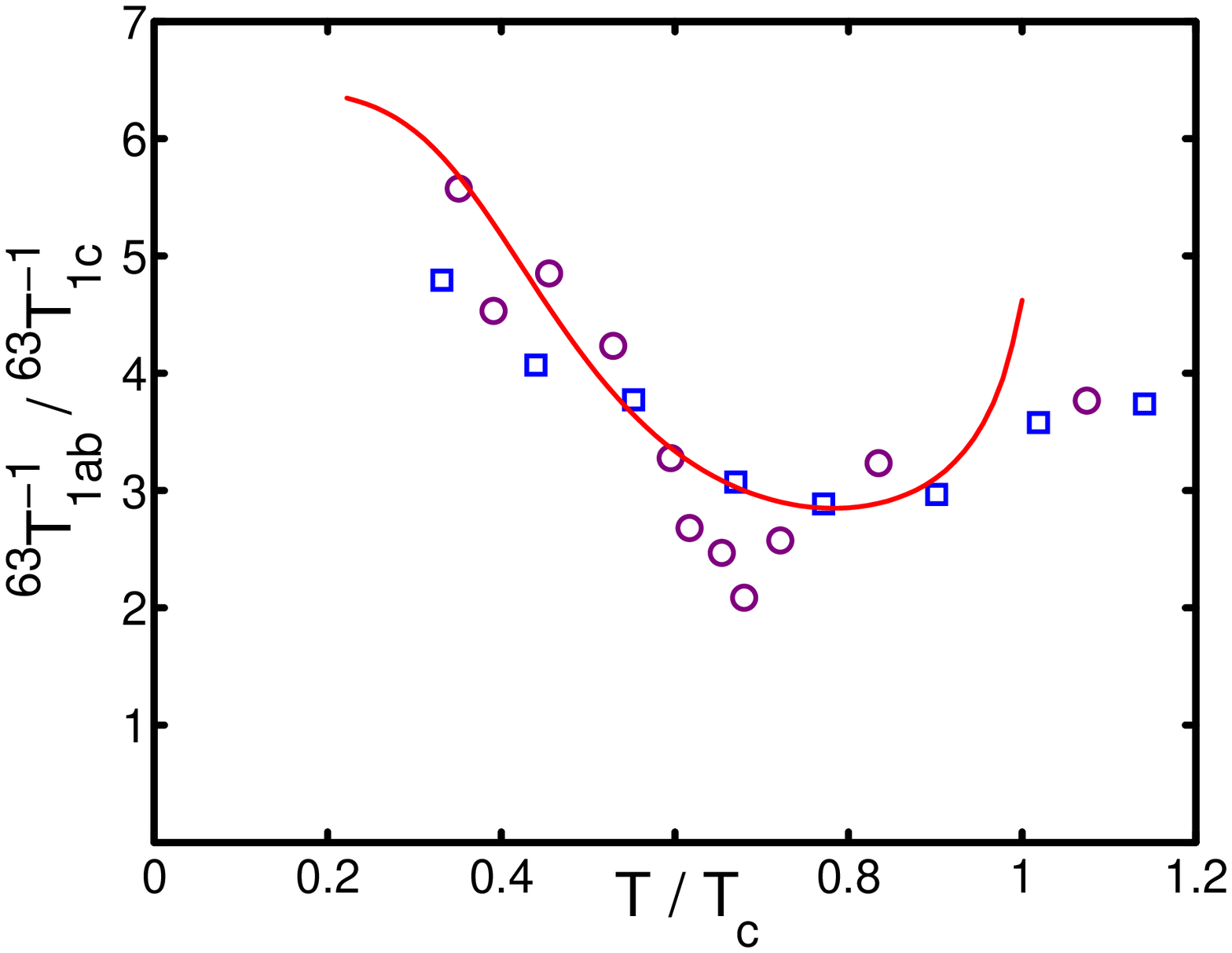}\end{center}

\begin{center}\includegraphics[%
  scale=0.4]{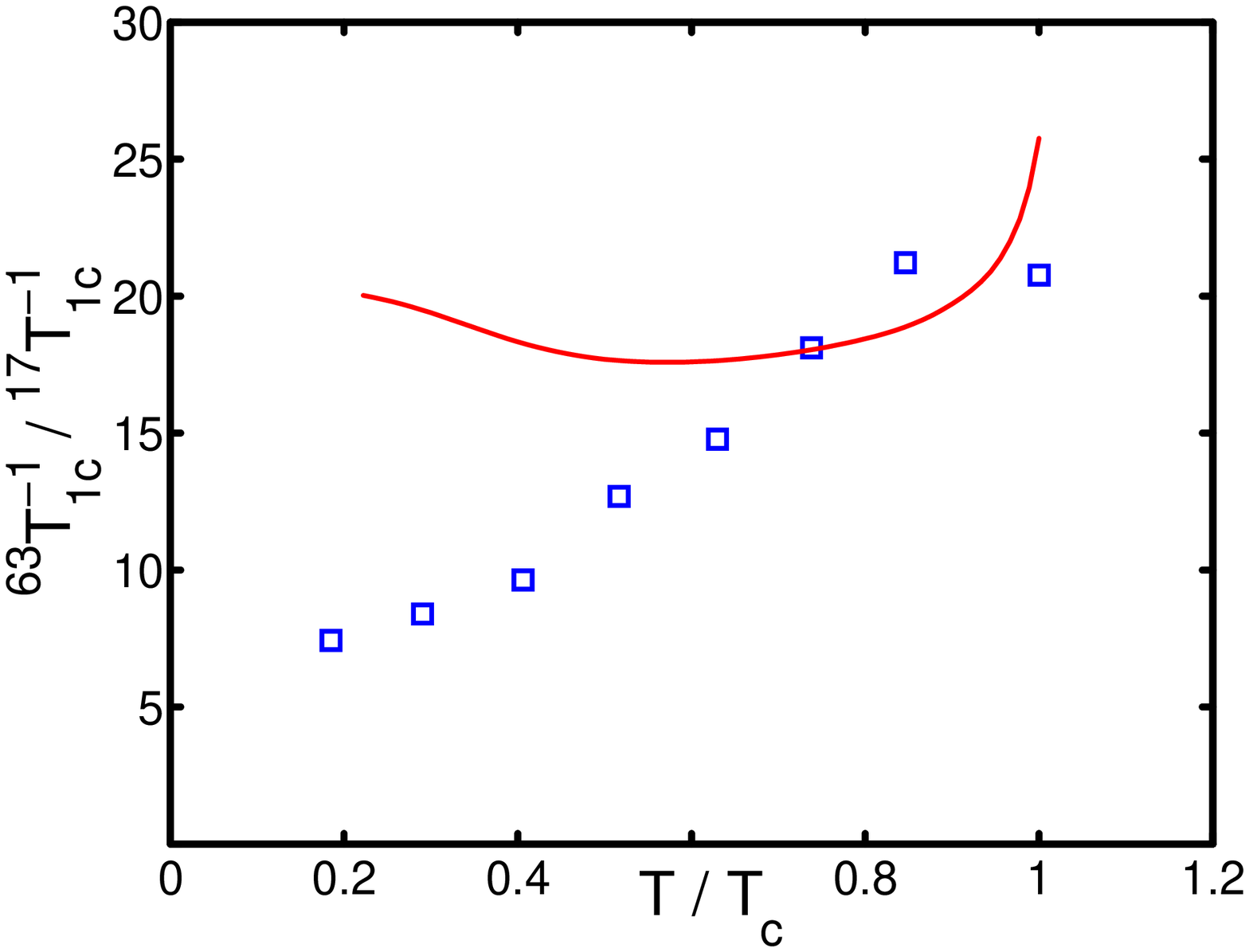}\end{center}

\caption{\label{cap:ani}Temperature dependence of the anisotropy ratios $^{63}T_{1c}^{-1}/^{17}T_{1c}^{-1}$
and $^{63}T_{1ab}^{-1}/^{63}T_{1c}^{-1}$ in YBa$_{2}$Cu$_{3}$O$_{7}$
(squares). Data for YBa$_{2}$Cu$_{4}$O$_{8}$ are also shown for
comparison (circles). }
\end{figure}

We are also interested in the anisotropy ratios $^{63}T_{1ab}^{-1}/^{63}T_{1c}^{-1}$
and $^{63}T_{1c}^{-1}/^{17}T_{1c}^{-1}$ measured in YBa$_{2}$Cu$_{3}$O$_{7}$.
Experimental evidence\cite{martindale ani,takigawa ani,bankay ani}
points toward a field dependence of these quantities. Our theoretical
results have to be compared with data in zero or small external fields.
We display the calculated anisotropy ratios in Fig.\ \ref{cap:ani}.
The experimental points for $^{63}T_{1c}^{-1}/^{17}T_{1c}^{-1}$ are
from Martindale et al.\cite{martindale ani}, whereas those for $^{63}T_{1ab}^{-1}/^{63}T_{1c}^{-1}$
are taken from Takigawa et al.\cite{takigawa ani} (squares). %
 For comparison we also plotted the anisotropy in the YBa$_{2}$Cu$_{4}$O$_{8}$
compound (circles) measured by Bankay et al.\cite{bankay ani}. For
the calculation we used the same parameters as before, $\Delta_{0}=22$~meV
and $J_{1}=90$~meV. We see that we can account for the anisotropy
ratio $^{63}T_{1ab}^{-1}/^{63}T_{1c}^{-1}$, but it is not possible
to reproduce the temperature dependence of the ratio $^{63}T_{1c}^{-1}/^{17}T_{1c}^{-1}$.
Our results for the weak--field anisotropy ratio $^{63}T_{1ab}^{-1}/^{63}T_{1c}^{-1}$
for YBa$_{2}$Cu$_{3}$O$_{7}$ agree with the weak--coupling calculations
of Bulut and Scalapino\cite{bulut-scalapino5}, which are based on
a square Fermi surface with nearest neighbour hopping only. However,
our calculations disagree with the analysis of Mack et al.\cite{mack},
where the Fermi surface for YBa$_{2}$Cu$_{3}$O$_{7}$ was assumed
to be quite different (similar to that which we used here to describe
the Bi$_{2}$Sr$_{2}$CaCu$_{2}$O$_{8}$ compound). Therefore we
conclude that the Fermi surface topology plays an important role in
the description of the weak--field anisotropy ratio $^{63}T_{1ab}^{-1}/^{63}T_{1c}^{-1}$.
Note that the experimentally reported weak--field ratio $^{63}T_{1c}^{-1}/{}^{17}T_{1c}^{-1}$
could also not be reproduced within previous weak--coupling RPA calculations\cite{bulut-scalapino5,mack}.

\section{Conclusions\label{sec:Conclusions}}

In summary we have determined the spin susceptibility in cuprates
within a special Hubbard model which includes strong correlation effects.
It has been found that the susceptibility in the strong--coupling
limit is different from the standard Pauli--Lindhard formula. In particular,
two correction functions were determined in the superconducting state.
The first one, $\Pi(\mathbf{q},\omega)$, found originally by Hubbard
and Jain\cite{hubbard-jain} in the normal state, originates from
the anticommutator rule which is modified due to the Coulomb repulsion,
whereas the function $Z(\mathbf{q},\omega)$ has its origin in the
fast fluctuations of the localized spins and was previously discussed
by Zavidonov and Brinkmann\cite{zavidonov-brinkmann} for the normal
state. 

We analyzed inelastic neutron scattering and NMR data in the superconducting
state of the optimally doped high--T$_{c}$ superconductors YBa$_{2}$Cu$_{3}$O$_{7}$
and Bi$_{2}$Sr$_{2}$CaCu$_{2}$O$_{8}$. In our analysis we have
taken into account the experimentally measured topology of the Fermi
surface, which is quite different for these two materials. We found
that on the whole the results within the strong--coupling and weak--coupling
limits agree with each other. Based on the results of our numerical
calculations we conclude that strong correlation effects, i.e. the
effect of the functions $\Pi(\mathbf{q},\omega)$, $Z(\mathbf{q},\omega)$
on the susceptibility can be modeled in the weak--coupling approach
by an appropriate redefinition\cite{eremin-manske} of the effective
Coulomb interaction parameter $U_{e}$. In particular, the non--physical
value of $U_{e}$ in the weak--coupling limit (sometimes $U_{e}\geq t$)
becomes understandable. In terms of our model this can be explained
by the impact of the $\Pi(\mathbf{q},\omega)$ function, which is
of the order of $t\chi_{0}^{+,-}(\mathbf{q},\omega)$. As concerns
the $Z(\mathbf{q},\omega)$ function, our calculations show that if
it is not included, the numerical values of $Im\chi^{+,-}(\mathbf{q},\omega)$
can for some set of parameters become negative, implying the importance
of this correction.

In the framework of the singlet--correlated band model we found it
possible to describe the available experimental data in the optimally
doped YBa$_{2}$Cu$_{3}$O$_{7}$ and Bi$_{2}$Sr$_{2}$CaCu$_{2}$O$_{8}$
compounds within one set of model parameters for each material. These
optimal sets of parameters are given by $\Delta_{0}=23$~meV ($\pm5$\%),
$J_{1}=90$~meV for YBa$_{2}$Cu$_{3}$O$_{7}$ and $\Delta_{0}=24$~meV
($\pm5$\%), $J_{1}=110$~meV for Bi$_{2}$Sr$_{2}$CaCu$_{2}$O$_{8}$.
The only experiment which could not be reproduced is the temperature
dependence of the weak--field anisotropy ratio $^{63}T_{1c}^{-1}/^{17}T_{1c}^{-1}$
in YBa$_{2}$Cu$_{3}$O$_{7}$. As concerns the Bi$_{2}$Sr$_{2}$CaCu$_{2}$O$_{8}$
material, more experimental data would be desirable to test our model
further, for example low energy inelastic neutron scattering and spin--spin
relaxation rate measurements.

\begin{acknowledgments}
This work is supported by the Swiss National Science Foundation. M.
Eremin was partially supported by the Russian RFFI (grant No. 03 02
17430) and Superconductivity program. We would  like to thank M. Mali
and J. Roos for fruitful discussions.
\end{acknowledgments}

\end{document}